\documentclass[12pt]{iopart}
\usepackage{graphicx,subfigure}
\usepackage{textcomp}
\begin{document}

\title{Domain-Wall Pinning by Local Control of Anisotropy in Pt/Co/Pt strips}

\author{J.H. Franken, M. Hoeijmakers, R. Lavrijsen, H.J.M. Swagten}
\address{
Department of Applied Physics, center for NanoMaterials (cNM),
Eindhoven University of Technology, P.O.Box 513, 5600 MB
Eindhoven, The Netherlands
}
\ead{j.h.franken@tue.nl}

\begin{abstract}
We theoretically and experimentally analyze the pinning of a magnetic domain wall (DW) at engineered anisotropy variations in Pt/Co/Pt strips with perpendicular magnetic anisotropy. An analytical model is derived showing that a step in the anisotropy acts as an energy barrier for the DW. Quantitative measurements are performed showing that the anisotropy can be controlled by focused ion beam irradiation with Ga ions. This tool is used to experimentally study the field-induced switching of nanostrips which are locally irradiated. The boundary of the irradiated area indeed acts as a pinning barrier for the domain wall and the pinning strength increases with the anisotropy difference. Varying the thickness of the Co layer provides an additional way to tune the anisotropy, and it is shown that a thinner Co layer gives a higher starting anisotropy thereby allowing tunable DW pinning in a wider range of fields. Finally, we demonstrate that not only the anisotropy itself, but also the width of the anisotropy barrier can be tuned on the length scale of the domain wall.
\end{abstract}

\maketitle

\section{Introduction}
The ability to propagate a domain wall (DW) through a submicron magnetic wire using a magnetic field \cite{Ono1999,Allwood2002,Nakatani2003,Atkinson2003,Beach2005} or electric current \cite{Vernier2004,Yamaguchi2004,KLAUI2005,Meier2007,BEACH08} is the basis of several new spintronics devices \cite{Allwood2005,Parkin2008,HAYASHI08,Zutic2004,Xu2008}. Regarding the topic of current-induced DW dynamics, most is known about DWs in in-plane magnetized permalloy strips \cite{KLAUI08}. Recently, the focus has been shifting toward materials with high perpendicular magnetic anisotropy (PMA) \cite{RAVELOSONA2005,Boulle2008a,Moore2008,SanEmeterioAlvarez2010,Burrowes2009,Miron2011,KIM10,Koyama2011,Heinen2010}. Although field-driven DW motion is typically slow due to DW creep \cite{LEMERLE1998,METAXAS2007,KIM2010}, these materials might show faster current-induced DW motion, because they exhibit simple and narrow DWs potentially leading to large non-adiabatic spin torque contributions \cite{Thiaville2005,Zhang2004,Tatara2004}, or by the presence of Rashba fields stabilizing the DW structure during propagation \cite{Miron2011}. Furthermore, recent results indicate that the non-adiabaticity is strongly dependent on details of the perpendicular material, ranging from a negligible effect in Co/Ni \cite{Koyama2011} to a large contribution in Pt/Co multilayers \cite{Boulle2008a,Heinen2010}. These interesting observations call for more experiments on various material systems.

Being able to control the position of DWs at will is essential for successful DW experiments or devices. One issue is the initial creation of a magnetic domain and its domain walls. A second issue is to control the exact pinning positions where a domain wall stops after propagation, which is needed in several memory and logic devices making use of spintronics \cite{Parkin2008,Zutic2004,Xu2008}. For the first issue of writing a domain at a controlled position, there are generally two possibilities: one should either apply a highly localized magnetic field, or locally modify the switching properties of the magnetic nanostrip to be able to write with a global field. A highly localized magnetic field poses restrictions to the experimental environment and therefore writing with a global field is often the desired option. For in-plane magnetized DW devices made of permalloy, one often designs a variation in shape, such as a bend in the wire \cite{Allwood2002,KLAUI2005} or a large pad at the end of the wire \cite{Atkinson2003,Shigeto1999,Thomas2005}. Due to shape anisotropy, these lead to preferential nucleation points when an external field is applied. For PMA materials however, there is a very strong perpendicular easy axis that dominates over shape-induced effects, by which nucleation preferably occurs at randomly distributed defects. For the second issue of controlled DW pinning, similar considerations apply in PMA materials: geometric variations can be used for DW pinning \cite{RAVELOSONA2005,Burrowes2009} but these shape-induced effects are rather weak and typically lead to deformations of the domain wall \cite{RAVELOSONA2005}, causing the DW to lose its one-dimensional (1D) character.

In a recent study \cite{JeroenFIB} it was shown that both issues can be tackled at the same time by taking control over the parameter that governs the switching behavior: the PMA. The PMA is known to be reduced by irradiation with highly energetic ions \cite{Fassbender2008,Hyndman1,Chappert1998,Devolder2000,Devolder2001,Vieu2002}. Using a focused ion beam (FIB) of, for example, Ga \cite{JeroenFIB,Hyndman2,Aziz2005,Aziz2006} or He \cite{Markie} ions, the anisotropy can be controlled very locally (at a scale of a few nanometer). By locally reducing the anisotropy, the coercivity is also reduced and a DW nucleation area is made. Furthermore, it was shown that DWs tend to pin at a discontinuity in the anisotropy, i.e. the boundary of a Ga-irradiated area, solving the second issue. In the current paper, we provide further insights into this pinning of DWs at engineered anisotropy variations. First, we describe in detail the mechanism responsible for DW pinning at anisotropy variations, through the development of a 1D model in section \ref{sec:model}. Furthermore, the magnetic anisotropy of Pt/Co/Pt strips is experimentally determined as a function of Ga irradiation dose and Co layer thickness in section \ref{sec:Anisotropy}. Finally, in section \ref{sec:DWpinning}, we report a detailed experimental study on DW pinning at an anisotropy boundary, showing that the DW energy landscape in a nanostrip can basically be engineered at will on a nanometer scale.

\section{Model of DW pinning}\label{sec:model}

\begin{figure}[htb]
     \begin{center}
      \includegraphics[width=0.7\linewidth]{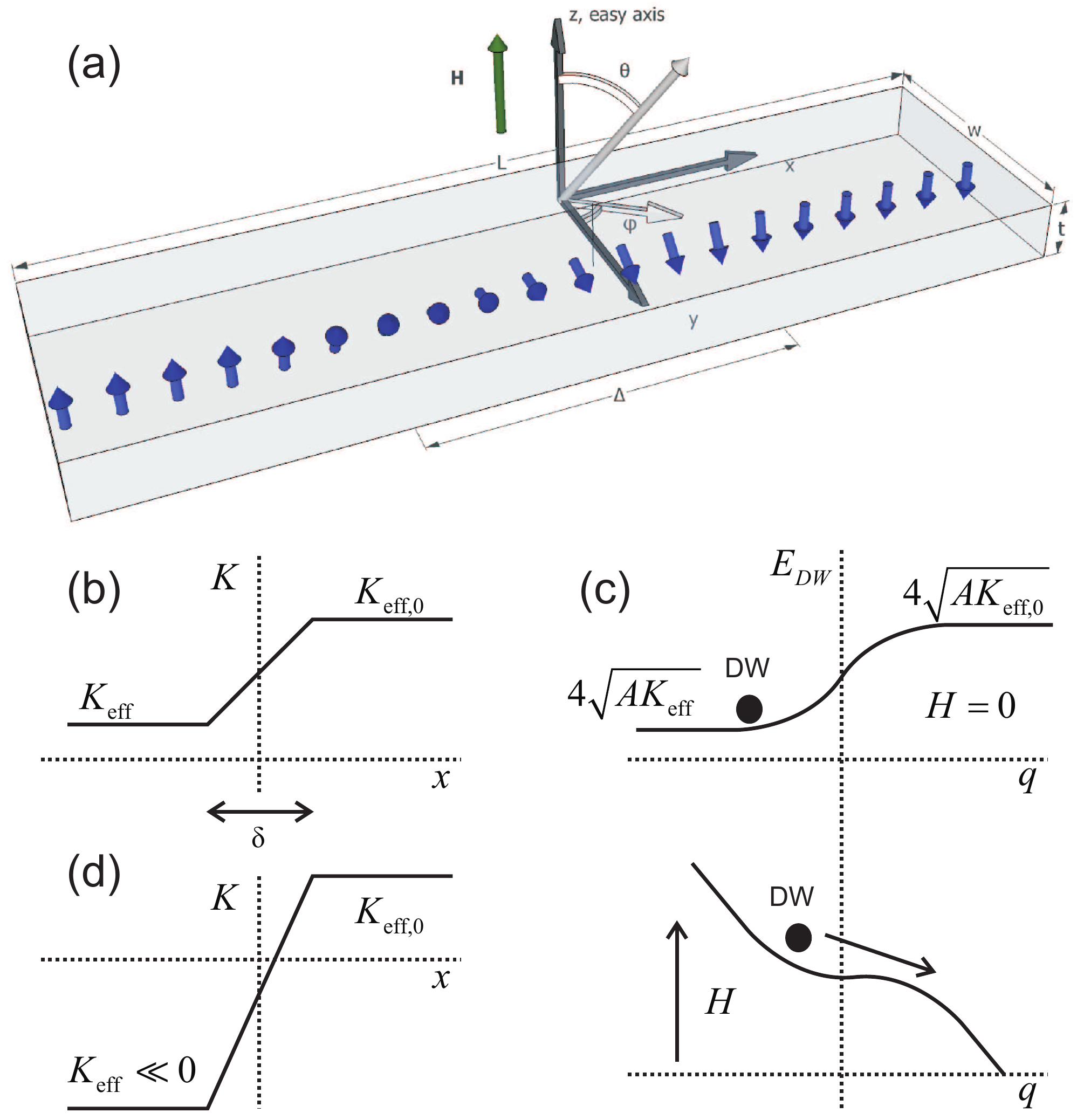}
      \caption{(a) Sketch of a Bloch DW in a nanostrip and definition of the coordinate system. (b) Sketch of a step in the anisotropy along the strip direction $x$. Such a step leads to an energy barrier for a DW sitting to the left of the step, as sketched in (c). The barrier can be overcome by applying an external magnetic field that tilts the energy landscape. (d) Sketch of the anisotropy landscape in case the part $x\ll0$ has in-plane magnetic anisotropy ($K_{\rm{eff}}<0$). }
      \label{Figure1}
    \end{center}
\end{figure}

In this section, we investigate how DWs are pinned at anisotropy modulations by assuming a simple model system (figure \ref{Figure1}(a)). The system consists of a PMA strip of length $L$, width $w$, and thickness $t$. We assume that a single 1D Bloch DW is present in the strip, at a certain position $q$ along the $x$-axis. The strip has perpendicular magnetic anisotropy, but the anisotropy changes at $x=0$. We assume a linear transition between two values over a gradient length $\delta$ centered at $x=0$. The part $x<- \delta/2$ has an effective perpendicular anisotropy constant $K_{\rm{eff}}$ and the part $x>\delta/2$ has $K_{\rm{eff,0}} >K_{\rm{eff}}$ (figure 1(b)).  The other relevant parameters $M_{\rm{s}}$ (saturation magnetization) and $A$ (exchange constant) are kept constant.  Since the energy of a DW scales with the square root of the anisotropy, the anisotropy change at $x=0$ causes an energy barrier as sketched in figure 1(c). The larger the anisotropy difference, the larger this barrier. By applying an external field $H$, the potential landscape is tilted making it possible for the DW to escape as soon as the tilt slope cancels the maximum slope of the DW energy landscape.

In the following, we derive expressions for the pinning field $H_{\rm{pin}}$ as a function of the anisotropy of the left part of the strip, $K_{\rm{eff}}$. We will discuss the two cases shown in figure \ref{Figure1}(b) and \ref{Figure1}(d). The situation of figure \ref{Figure1}(b), in the limit that the anisotropy step is small, is discussed in section \ref{sec:modelSmallStep}. In section \ref{sec:ModelIP}, we discuss the situation where the part $x<0$ has strong in-plane (shape) anisotropy, $K_{\rm{eff}} \ll 0$, as sketched in figure \ref{Figure1}(d). We compare the analytical model with full micromagnetic simulations and find exact agreement.

\subsection{Limit of small $K$ step} \label{sec:modelSmallStep}
% TODO: is there a relation to work from the 60s?

A DW centered at a position $q$ in a perpendicularly magnetized nanostrip has a standard Bloch profile, with the out-of-plane angle $\theta$ given by \cite{SLONCZEWSKI79}
\begin{equation} \label{e:gallium:bloch}
\theta(x) = \pm 2 \arctan \left[\exp\left(\frac{x-q}{\Delta}\right)\right],
\end{equation}
with $\Delta$ the DW width. This profile is not exactly valid in the vicinity of the anisotropy interface since the part of the DW residing in the low-$K$ region tends to widen, but this effect is negligible in the limit studied. Considering effective anisotropy and exchange contributions, the magnetic energy density is \cite{SLONCZEWSKI79}
\begin{eqnarray}\label{energydens}
w(x)&=A\left[\left(\frac{\partial \theta}{\partial x}\right)^2 + \left(\sin \theta \frac{\partial \phi}{\partial x}\right)^2 \right] + K(x) \sin^2 \theta \nonumber \\
&= \left(\frac{A}{\Delta^2} + K(x) \right) \mbox{sech}^2 \left(\frac{x-q}{\Delta}\right),
\end{eqnarray}
where $\phi$ is the in-plane angle of magnetization ($\phi=0$ for a Bloch DW), and $K(x)$ has the profile sketched in figure \ref{Figure1}(b),
\begin{eqnarray}
K (x) =&  K_{\rm{eff}}& (x<-\delta/2), \nonumber\\
K (x) =&  \frac{K_{\rm{eff,0}} + K_{\rm{eff}}}{2} + (K_{\rm{eff,0}} - K_{\rm{eff}})\frac{x}{\Delta} \qquad & (-\delta/2\leq x \leq \delta/2), \\
K (x) =&  K_{\rm{eff,0}}& (x>\delta/2). \nonumber
\end{eqnarray}
Because the DW width can be considered constant in the limit studied, the term $A/\Delta^2$ in (\ref{energydens}) can be omitted for simplicity. The total DW energy per unit cross-sectional area $\sigma_{\rm{DW}}$ of a DW centered at $q$ is then (up to constant) given by
\begin{eqnarray} \label{e:gallium:integral}
\sigma_{\rm{DW}} (q) = \int_{-\infty}^{\infty}  w(x) \mathrm{d}x =  &\frac{\Delta}{\delta }  \left(2 K_{\rm{eff,0}} \delta +(K_{\rm{eff,0}}-K_{\rm{eff}}) \,\Delta \right. \nonumber\\
& \left. \times  \left(\ln\left[1+e^{-\frac{2 q+\delta }{\Delta }}\right] -\ln\left[1+e^{\frac{-2 q+\delta }{\Delta }}\right]\right)\right).
\end{eqnarray}
By applying an external magnetic field $H$ in the $z$-direction, the energy landscape of the domain wall is tilted due to the Zeeman energy, giving a total energy $\varepsilon(q)$
\begin{equation}
\varepsilon(q) = \sigma_{\rm{DW}}(q)  -2\mu_0 M_{\rm{s}} H q .
\end{equation}

For estimating the depinning field, we are interested in the derivative of the DW energy with respect to $q$, which should be negative at any position in order for the DW to depin,
\begin{equation}\label{e:gallium:derivative}
\frac{\mathrm{d}\varepsilon}{\mathrm{d} q} = \frac{2 (K_{\rm{eff,0}}-K_{\rm{eff}}) \Delta \sinh\left[\frac{\delta }{\Delta }\right]}{\delta  \left(\cosh\left[\frac{2 q}{\Delta }\right]+\cosh\left[\frac{\delta }{\Delta }\right]\right)} - 2 \mu_0 M_{\rm{s}} H < 0\,..
\end{equation}
Hence, the maximum of $\frac{\mathrm{d}\varepsilon}{\mathrm{d} q}$ should be negative,
\begin{equation}
\max_{-\infty<q<\infty} \frac{\mathrm{d}\varepsilon}{\mathrm{d} q} =  \left.\frac{\mathrm{d}\varepsilon}{\mathrm{d} q}\right|_{q=0} = (K_{\rm{eff,0}} - K_{\rm{eff}} )\frac{2\Delta}{\delta}\tanh\frac{\delta}{2\Delta}- 2 \mu_0 M_{\rm{s}} H < 0.
\end{equation}
The DW thus depins for $H > H_{\rm{pin}}$, with
\begin{equation} \label{e:gallium:pinningfield}
H_{\rm{pin}} = \frac{K_{\rm{eff,0}} - K_{\rm{eff}} }{{2{\mu _0}{M_{\rm{s}}}}} \times \frac{{2\Delta }}{\delta }\tanh \frac{\delta }{2\Delta }.
\end{equation}
In case the length scale of the anisotropy gradient $\delta$ is much smaller than the DW width $\Delta$, the pinning field is simply given by the difference of the anisotropy values,
\begin{equation} \label{e:gallium:pinningfieldSharp}
\lim_{\delta \rightarrow 0} H_{\rm{pin}} = \frac{K_{\rm{eff,0}} - K_{\rm{eff}} }{{2{\mu _0}{M_{\rm{s}}}}}.
\end{equation}
The opposite limit is also interesting; it turns out that the pinning field becomes zero if $\delta \gg \Delta$,
\begin{equation} \label{e:gallium:pinningfieldSharp}
\lim_{\delta \rightarrow \infty } H_{\rm{pin}} = 0,
\end{equation}
which means that a DW will only pin if $\delta$ is at a length scale comparable to the DW width, typically in the range of $10$\,nm.

\subsection{Limit of in-plane $K$} \label{sec:ModelIP}

If the perpendicular uniaxial anisotropy is quenched completely, this results in an effective in-plane anisotropy. Therefore, the DW at the moment of depinning is not necessarily an `up' to `down' transition. If the effective in-plane anisotropy is small, the out-of-plane field that is applied to achieve DW injection is already enough to pull the magnetization fully out-of-plane and the origin of the DW pinning field is not physically different from the case studied in the previous section. However, if the in-plane shape anisotropy is strong, there will always be a 90$^{\circ}$ DW present at the interface, and reversal is merely initiated by nucleation of a DW at this interface that will propagate through the out-of-plane part of the strip. In the following, we will attempt to model this situation by assuming that the in-plane anisotropy is so large that the spins are completely in-plane in the irradiated area, even though a perpendicular field is applied. This in fact corresponds to infinite in-plane anisotropy. Furthermore, it is assumed that the Bloch profile is still valid, but rescaled from the domain $\theta \in [0,\pi]$ to $\theta \in [0,\frac{\pi}{2}]$. The profile then reads (notice the factor 2 difference with (\ref{e:gallium:bloch}))
\begin{equation}
\theta(x) = \pm \arctan \left[\exp\left(\frac{x-q}{\Delta}\right)\right].
\end{equation}
By micromagnetic simulations of an in-plane to out-of-plane transition in a strip, we verified that this profile is reasonably precise. To simplify the calculation, we only consider the case $\delta = 0$, because the precise shape of the anisotropy profile was found not to matter in the limit studied. The DW energy density reflects the change of easy axis at $x>0$:
\begin{eqnarray}
w(x) &= \frac{A}{4\Delta^2} &  + \left| K_{\rm{eff}} \right| \cos^2 \theta \nonumber\\
& = \frac{A}{4\Delta^2} & + \left| K_{\rm{eff}} \right| \frac{1}{\exp\left(2\frac{x-q}{\Delta}\right)+1}  \qquad \mbox{ ($x<0$),}\\
w(x) &=  \frac{A}{4\Delta^2} & + K_{\rm{eff ,0}} \sin^2 \theta \nonumber \\
& = \frac{A}{4\Delta^2}  & + K_{\rm{eff,0}} \frac{\exp\left(2\frac{x-q}{\Delta}\right)}{\exp\left(2\frac{x-q}{\Delta}\right) +1} \qquad \mbox{ ($x>0$).}
\end{eqnarray}
In analogy with (\ref{e:gallium:derivative}), the derivative of $\sigma_{\rm{DW}}$ becomes
\begin{equation}
\frac{\mathrm{d}\sigma_{\rm{DW}}}{\mathrm{d} q} = \frac{K_{\rm{eff,0}}\exp\left(\frac{2q}{\Delta}\right) - \left|K_{\rm{eff}}\right|}{\exp\left(\frac{2q}{\Delta}\right) +1}.
\end{equation}
This function is monotonically increasing and is maximal at $q\rightarrow\infty$. Therefore, the maximum slope of the energy barrier is given by
\begin{equation}
\max_{-\infty<q<\infty} \frac{\mathrm{d}\sigma_{\rm{DW}}}{\mathrm{d} q}  = K_{\rm{eff,0}}.
\end{equation}
A more detailed analysis shows that at finite in-plane anisotropy, if a small $z$-component of magnetization is assumed for $x<0$, the maximum derivative is not at $\infty$ but close to $q=0$ (retaining the same magnitude), so that injection indeed occurs at the anisotropy interface. The derivative of total energy includes again a Zeeman term, which now has half the original magnitude, because the z-component of magnetization is zero at one end of the DW. Therefore,
\begin{equation} \label{e:gallium:lowlimitenergy}
\max_{-\infty<q<\infty} \frac{\mathrm{d}\varepsilon}{\mathrm{d}q} = K_{\rm{eff,0}}- \mu_0 M_{\rm{s}} H,
\end{equation}
and the pinning field is found by equating this expression to zero,
\begin{equation} \label{e:gallium:pinningfieldIP}
H_{\rm{pin}} = \frac{K_{\rm{eff,0}}}{\mu_0 M_{\rm{s}}}.
\end{equation}

\begin{figure}[tbh]
     \begin{center}
      \includegraphics[width=0.7\linewidth]{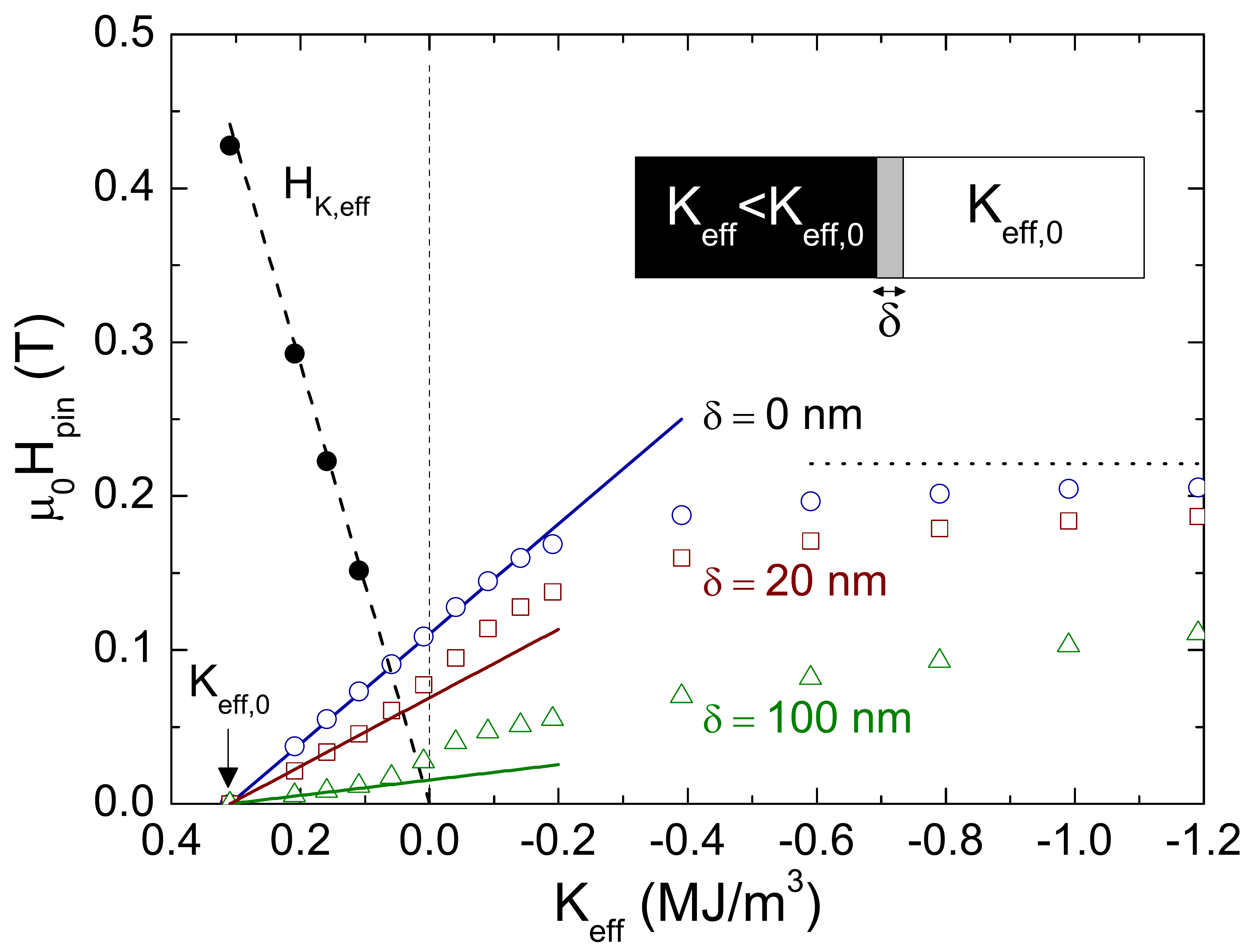}
      \caption{H$_{\rm{pin}}$ obtained from micromagnetic simulations of DW depinning at a sharp anisotropy step (open circles), or a gradual anisotropy increase (open squares and triangles). The solid and dotted lines show the limiting cases of the 1D model derived in the text. The filled circles are simulated nucleation fields of the left area, which dominate the switching of the entire strip if if reversal is started from a saturated state (as in experiment). Part of the data adapted from \cite{JeroenFIB,Markie}.
      \label{Figure2}}
    \end{center}
\end{figure}

To test the validity of (\ref{e:gallium:pinningfield}) and (\ref{e:gallium:pinningfieldIP}), micromagnetic simulations \cite{llg} are performed on a strip with $w=60\,$nm, $t=1$\,nm, and length $L=400\,$nm. The simulation cell size is $4\times4\times1$\,nm$^3$. Reducing the simulation cell size did not significantly change the obtained results. The saturation magnetization $M_{\rm{s}}=1400$\,kA/m and the exchange constant $A=16\,$pJ/m. The uniaxial anisotropy constant of the right part of the strip was fixed at $K_0 = 1.5$\,MJ/m$^3$,  yielding an effective anisotropy $K_{\rm{eff,0}} = K_0 - \frac{1}{2}\mu_0 N_z M_{\rm{s}}^2 = 0.305$\,MJ/m$^3$. The left part of the strip has a variable effective anisotropy $K_{\rm{eff}} < K_{\rm{eff,0}}$. The starting configuration is a DW that is artificially created at the boundary and then energetically relaxed at zero applied field. Then, the field is increased in small steps, and at each field step the LLG solver iterates until the torque on the magnetization is virtually zero. The result is shown in figure \ref{Figure2}.

The situation $\delta \rightarrow 0$ is shown as open circles in figure \ref{Figure2}, and $H_{\rm{pin}}$ from the 1D model (\ref{e:gallium:pinningfieldSharp}) is plotted as a solid line. In the regime where the anisotropy difference is rather small, good agreement is found. We also see that as the anisotropy becomes negative (in-plane), the simulated data approaches the derived limit (\ref{e:gallium:pinningfieldIP}), shown as the dotted horizontal line. The situation of a finite length $\delta$ is also simulated, by changing the values of the anisotropy on a single cell level. For instance, to simulate a length $\delta=20\,$nm, the anisotropy is step-wise increased over a width of 5 cells, which are each 4\,nm wide. This approach is valid as long as the cell size is of the order of the exchange length. Plotting the 1D limit (\ref{e:gallium:pinningfield}) in this case is slightly more complicated because it also contains the DW width $\Delta$, which in turn depends on the anisotropy at the DW position. For the plotted lines, we simply used $\Delta = \sqrt{A/K_{\rm{eff,0}}} \approx 7\,$nm, which again shows excellent agreement in the evaluated limit. Interestingly, for larger anisotropy differences, we see that the pinning field in the simulations bends upwards from the limit. This is simply because $\Delta$ increases, as it is partially in a region with lower $K_{\rm{eff}}$. If we take into account this increasing $\Delta$, the 1D model also predicts this upturn, demonstrating the power of the 1D approach.

In the experimental situation, starting from a saturated state, a DW does not readily exist but must first be nucleated. Therefore, simulations starting from the saturated state were also conducted, shown as the solid circles in figure \ref{Figure2}. It is consistently observed that the DW is nucleated in the left part of the strip. For relatively high $K_{\rm{eff}}$, the nucleation field is much higher than the pinning field and therefore dominates the switching field of the entire strip. The nucleation field in the simulations matches that of a Stoner-Wohlfart particle and is in good approximation given by the anisotropy field $H_{K_{\rm{eff}}} = 2 K_{\rm{eff}}/(\mu_0 M_{\rm{s}})$, plotted as the dashed line. We should note that this nucleation field has no quantitative meaning in experiments, where the switching behavior does not show coherent Stoner-Wohlfart behavior, but is dominated by domains nucleating at random defects and their expansion by DW motion.

To conclude this section, we have shown by analytical modeling and micromagnetic simulations, that a DW can be pinned at an anisotropy boundary. The field strength needed for depinning depends linearly on the anisotropy difference if the boundary is not too high. Interestingly, it was shown that a DW can also be injected from a boundary between an in-plane and out-of-plane anisotropy region. Furthermore, not only the height of the anisotropy boundary, but also its spatial extent (width), is an extra parameter that tunes the pinning field, and should be at the length scale of the DW for pinning to occur. In the next sections, we study quantitatively how Ga FIB irradiation can be used to tune the anisotropy (Section \ref{sec:Anisotropy}), and how DW pinning and nucleation can be controlled using this tool (Section \ref{sec:DWpinning}).

\section{Manipulating the anisotropy of Pt/Co/Pt}
\label{sec:Anisotropy}

Whereas it is widely accepted that Ga and He irradiation reduces the PMA of sputtered Pt/Co/Pt films, the evidence is usually indirect, i.e. through measurement of the coercive field. The anisotropy has been systematically measured as a function of He irradiation dose \cite{Devolder2000}, but to our knowledge, a systematic data set of anisotropy as a function of Ga dose is lacking. Performing a quantitative measurement of the anisotropy as a function of Ga dose is therefore interesting in its own right, as well as insightful for the interpretation of DW pinning and nucleation in section \ref{sec:DWpinning}.

Common methods to quantitatively measure the anisotropy of magnetic samples make use of Stoner-Wohlfart theory \cite{STONER1948}. Typically, an external field $H$ is applied under an angle $\alpha$ with the easy axis of magnetization. The magnetization is pulled away from its favored direction, toward the field direction. The ease by which the magnetization can be pulled is a measure of the anisotropy. We use the Extraordinary Hall Effect (EHE) to measure  $M_z(H,\alpha)$ on Hall crosses that have been irradiated with varying Ga doses, and obtain quantitative values for $K_{\rm{eff}}$ by fitting to the theoretical model \cite{Rosenblatt2010}.

\subsection{Experimental Details}

\begin{figure}[htbp]
	\centering
		\includegraphics[width=0.7\linewidth]{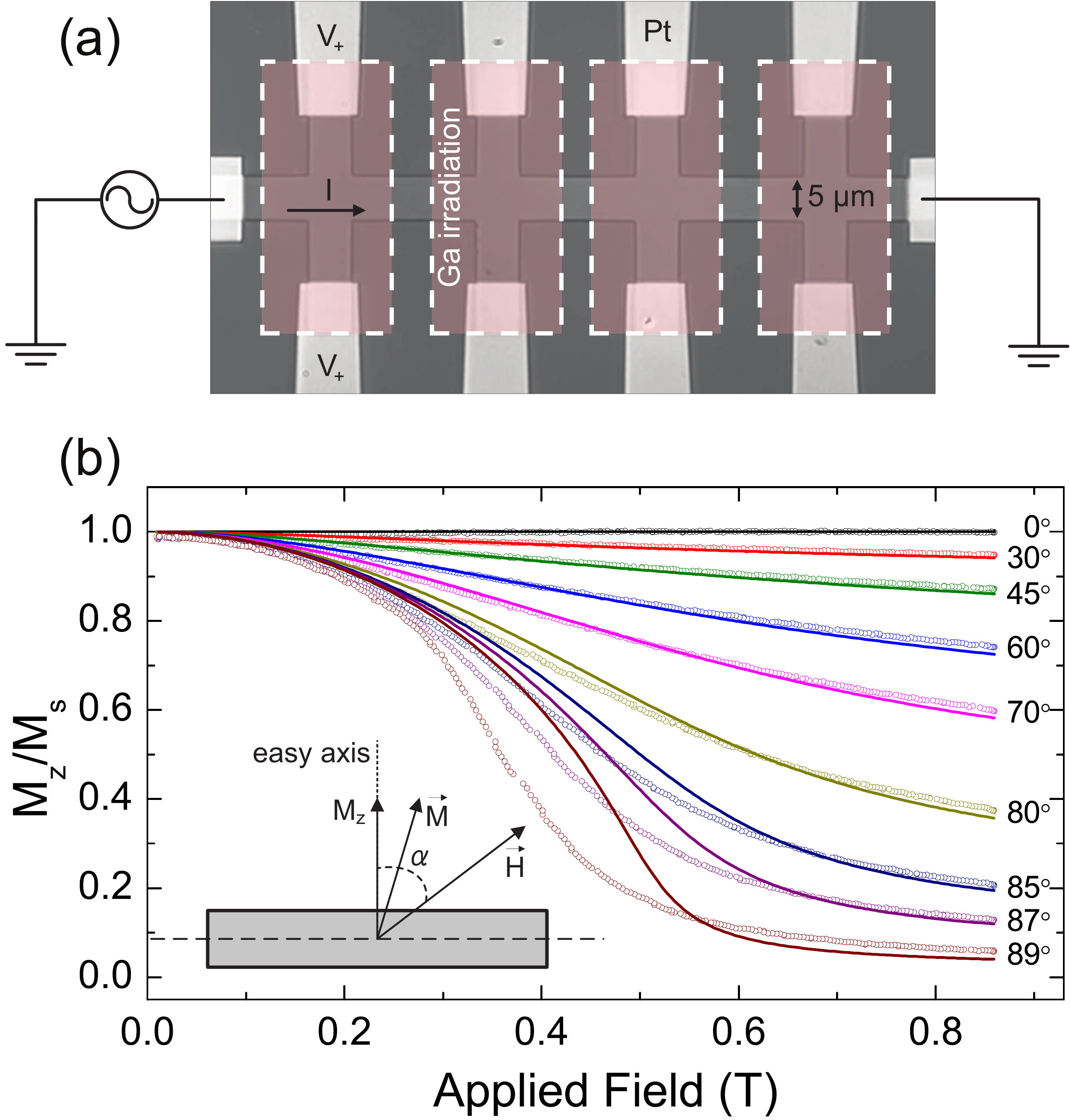}
	\caption{(a) Pt/Co/Pt sample with four irradiated Hall crosses for EHE measurements; (b) Example of $M_z(H,\alpha)$ (open circles). The lines are the result of a global Stoner-Wohlfart fit for all $\alpha$ up to 80$^{\circ}$. Higher $\alpha$ are not incorporated because of non-coherent magnetization reversal \cite{Rosenblatt2010}. The inset shows the experimental geometry.}
	\label{Figure3}
\end{figure}

Samples containing four Hall crosses of 5\,$\hbox{\textmu}$m wide Pt(4\,nm) / Co($x$\,nm) / Pt(2\,nm) are deposited on a  Si / SiO$_2$(100\,nm) substrate. The thickness of the Co layer is varied from 0.4 to 0.6 nm. The samples were fabricated using Electron Beam Lithography (EBL), sputtering and lift-off. On top of the branches of the Hall crosses, 20 nm thick Pt contacts are deposited using a second EBL step for electrical contact. A micrograph of the resulting sample is shown in figure \ref{Figure3}(a).

After the deposition of the Pt contacts, the Hall crosses are irradiated with different Ga doses. The ions have an energy of 30 keV and a beam current of several pA is used. The dose is varied from $0.07 \times 10^{13}\,$ions/cm$^{2}$ to $1.3 \times 10^{13}\,$ions/cm$^{2}$. This dose range does not lead to significant etching, but only affects the Pt/Co interfaces \cite{Hyndman1,Hyndman2}. The irradiated region for each Hall cross is indicated in figure \ref{Figure3}(a).

Four lock-in amplifiers are used to measure the EHE as a function of applied magnetic field on four different Ga-irradiated crosses at the same time. An AC current with a density of $\sim 3.0 \cdot 10^{9}\,$Am$^{-2}$ at a frequency of 5 kHz is sent through the strip. The external field is applied under a variable angle $\alpha$. The measured lock-in voltage consists of the EHE plus a small contribution of the ordinary Hall effect (OHE). Since the EHE is constant when the magnetization is saturated, we can use the measured signal slope at high perpendicular fields to subtract the OHE from all other measurements.

Figure\,\ref{Figure3}(b) shows a typical measurement of $M_{z}/M_{\rm{s}}$ for various $\alpha$. All traces are fitted globally using a fitting routine based on energy minimization of the Stoner-Wohlfart model. Input parameters within the model are the applied field $H$, the angle $\alpha$, the perpendicular magnetization $M_{z}$ and the saturation magnetization $M_{\rm{s}}$. The latter is estimated at $1.4 \times 10^ {6}\,$A/m from SQUID measurements. The fit yields a value of the perpendicular anisotropy $K_{\rm{eff}}$. The second order crystalline anisotropy is found to be negligible and therefore is not taken into account in the final fit.

It can be seen in figure \ref{Figure3}(b) that for nearly in plane fields ($\alpha > 80^{\circ}$) there is a strong deviation between the fits and the experimental data. This is known to arise from non-coherent magnetization reversal processes, wherein the structure no longer behaves as a single magnetic domain \cite{Rosenblatt2010,HUBERT98}. To exclude this effect only measurements up to an angle of 80\,$^{\circ}$ are incorporated in the fit.

\subsection{Anisotropy of Ga irradiated Pt/Co/Pt}
\label{sec:EHEResults}

\begin{figure}[htbp]
	\centering
		\includegraphics[width=0.7\linewidth]{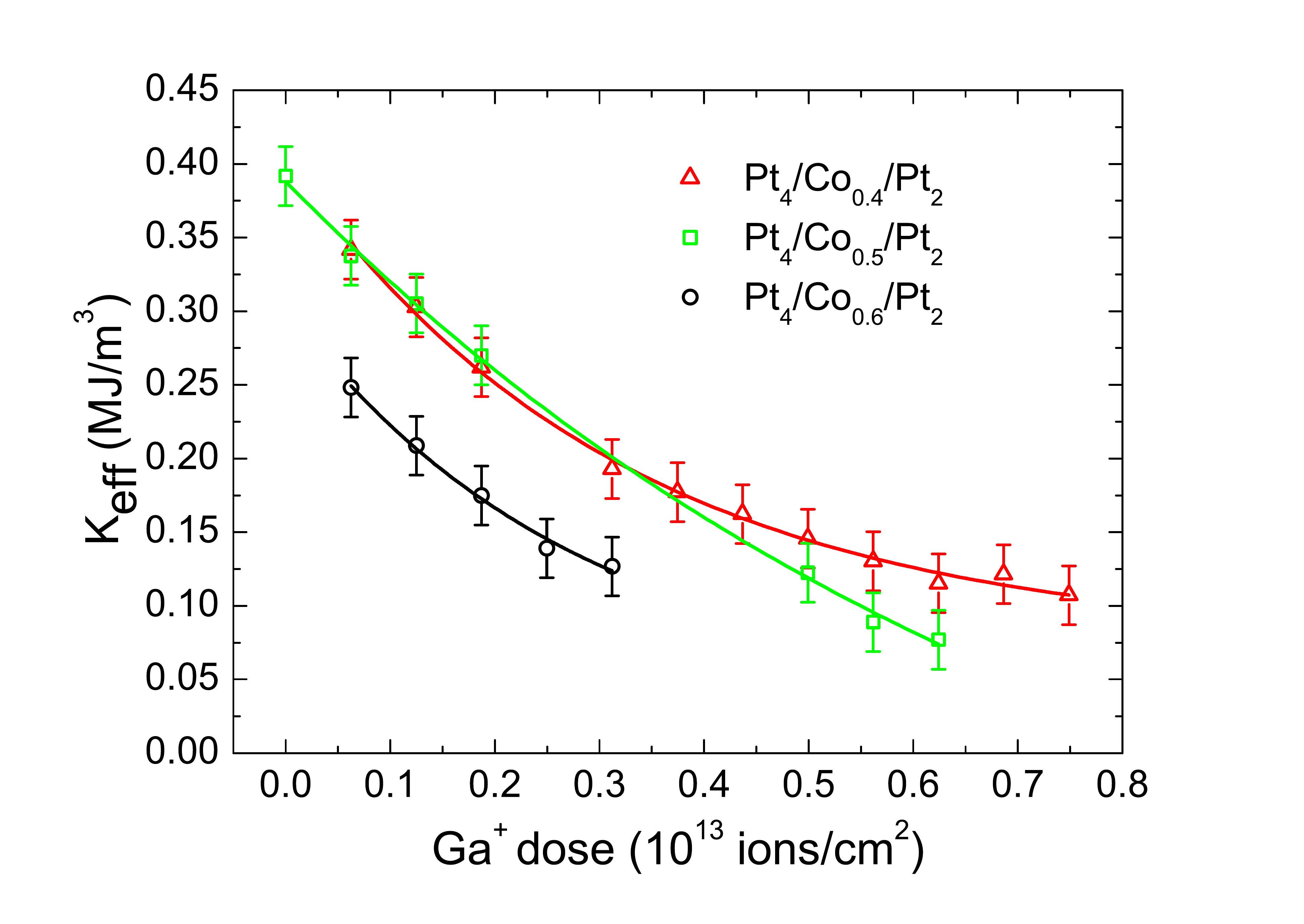}
	\caption{The anisotropy constant $K_{\rm{eff}}$ as a function of the Ga irradiation dose for Pt/Co/Pt structures with varying Co thickness. The guides are exponential fits.}
	\label{Figure4}
\end{figure}

Figure \ref{Figure4} visualizes the effect of Ga irradiation and Co layer thickness on the anisotropy of Pt/Co/Pt structures. First we discuss the influence of the Co layer. It is observed that the anisotropy increases if the Co thickness is reduced from 0.6 to 0.5 nm. This inverse dependence on $t$ is expected, since $K_{\rm{eff}}$ arises from the surface anisotropy $K_s$ at the Pt/Co interfaces via $K_{\rm{eff}} = 2 K_s/t +K_v$ \cite{JOHNSON96}, where $K_v$ is negative and contains the contribution from shape anisotropy.  However, the anisotropy of the 0.4\,nm Co sample does not differ significantly from the 0.5\,nm sample, meaning that growth-related phenomena are starting to play a role for such thin layers. Thinner layers are more ill-defined and therefore the interface anisotropy will decrease; this transition occurs right between 0.4 and 0.5\,nm. This is also reflected in a significantly lower coercivity of 0.4\,nm samples in section \ref{sec:DWpinning}, again pointing to a more disordered layer with easy nucleation centers.

As a function of Ga dose, we see a decrease of $K_{\rm{eff}}$ that is approximately linear at low dose, and less steep at high dose. For higher doses than shown, the remanence at zero field was significantly reduced and the Stoner-Wohlfart model could not be applied. Eventually, the magnetization becomes completely in-plane (negative  $K_{\rm{eff}}$). This transition to in-plane magnetization occurs at higher dose if the Co layer is thinner, because the anisotropy is higher to begin with. From a practical perspective this is very interesting, because the range of Ga doses that can be applied to tune the anisotropy increases by more than a factor of 2.

Whereas the effect of Ga irradiation on the anisotropy is now quantified, the effect on other magnetic properties is not. \emph{A priori}, however, we do not expect a very significant effect, since Ga irradiation mainly affects the interfaces and $M_{\rm{s}}$ and $A$ are typically bulk parameters. The magnitude of the EHE signal is some measure of $M_{\rm{s}}$, and we observed no trend as a function of Ga dose. Less is known about the effect on $A$, but at least such an effect is not needed for explaining the results in the remainder of this paper.

To conclude this section, it is seen that the anisotropy of Pt/Co/Pt samples increases for thinner Co layers, but this increase stops for very thin layers of $<0.5\,$nm. Interestingly, the reduction of anisotropy with low Ga dose remains constant irrespective of the starting anisotropy of the unirradiated film, i.e. the slope at low dose does not depend on the thickness in figure \ref{Figure4}. This is slightly counterintuitive, because if Ga irradiation reduces the surface anisotropy $K_s$ by the same amount regardless of thickness, this would translate to a $1/t$ dependence of the slope of $K_{\rm{eff}}$. From an experimental perspective this is a very useful result. By changing the Co thickness or the growth conditions, the tunable range of DW pinning fields can be expanded. In the next section \ref{sec:DWpinning} we will further investigate the consequences of this on the nucleation, pinning and injection of DWs in Pt/Co/Pt layers.

\section{Controlling Domain Wall Nucleation and Pinning}
\label{sec:DWpinning}

In the present section the effects of Ga irradiation on DW nucleation and pinning are investigated experimentally. First, the experimental method is described. In the subsequent sections, DW nucleation and pinning is investigated as a function of Ga dose, strip width, Co layer thickness, and beam focus. It will turn out that both the height and the width of the DW energy barrier can be tuned by these parameters.

\subsection{Experimental Details}

The investigated structures are rectangular Pt(4\,nm) / Co($x$\,nm) / Pt(2\,nm) strips of 15\,$\times$2\,$\hbox{\textmu}$m$^{2}$, 10\,$\times$1\,$\hbox{\textmu}$m$^{2}$, 5\,$\times$0.5\,$\hbox{\textmu}$m$^{2}$ and 2.5\,$\times$0.25\,$\hbox{\textmu}$m$^{2}$. Different Co thicknesses $x= 0.4$, 0.5 and 0.6\,nm are used. The structures are grown on a Si~/~SiO$_2$(100\,nm) substrate by EBL, sputtering, and lift-off.

After the fabrication of the Pt/Co/Pt layers, the left half of the strips is irradiated with Ga ions at a varying dose to reduce the anisotropy. Upon application of a magnetic field, a DW nucleates in this area and subsequently moves into the remainder of the strip. Wide-field Kerr microscopy \cite{HUBERT98} is used to study the effect of ion irradiation on nucleation and pinning of DWs.  In the analysis we focus on the injection field $H_{\rm{in}}$, defined as the external field at which the DW penetrates into the non-irradiated part of the structure. Since the injection of a DW involves two processes with a different typical field strength (nucleation at a field $H_{\rm{n}}$ and depinning at a field $H_{\rm{pin}}$), the injection field is defined as the maximum of these two fields. The magnetic field is swept from negative to positive and a sudden change in intensity of the Kerr signal occurs in the non-irradiated area when the DW is injected. Decent statistics are obtained by averaging $H_{\rm{in}}$ over 12 structures. The error bars in all figures where $H_{\rm{in}}$ is plotted against the irradiation dose represent the standard deviation of $H_{\rm{in}}$ from structure to structure.

\subsection{Variable Ga dose and strip width}

\begin{figure}[p]
   \begin{centering}
          \includegraphics[width=0.7\linewidth]{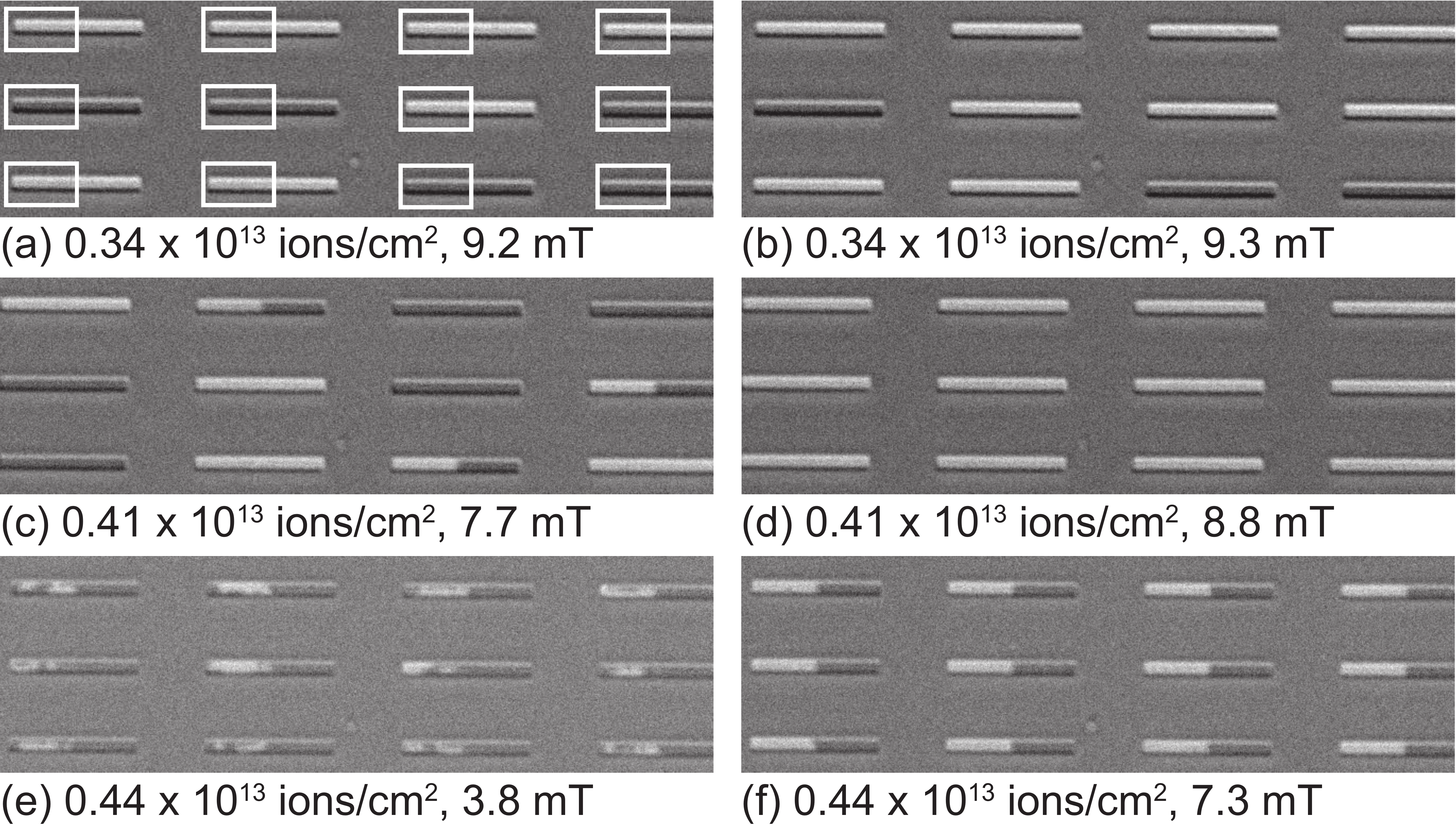}
      \caption{Kerr microscopy images of the magnetic switching behavior of 10\,$\times$1\,$\hbox{\textmu}$m$^{2}$ Pt / Co(0.6\,nm) / Pt structures for various doses of Ga irradiation. The irradiated regions are marked in (a). The magnetic contrast is enhanced by subtraction of a background image, which is obtained at zero field after saturation at high negative fields.}
      \label{Figure5}
          \end{centering}
\end{figure}

\begin{figure}[p]
   \begin{centering}
   \includegraphics[width=0.7\linewidth]{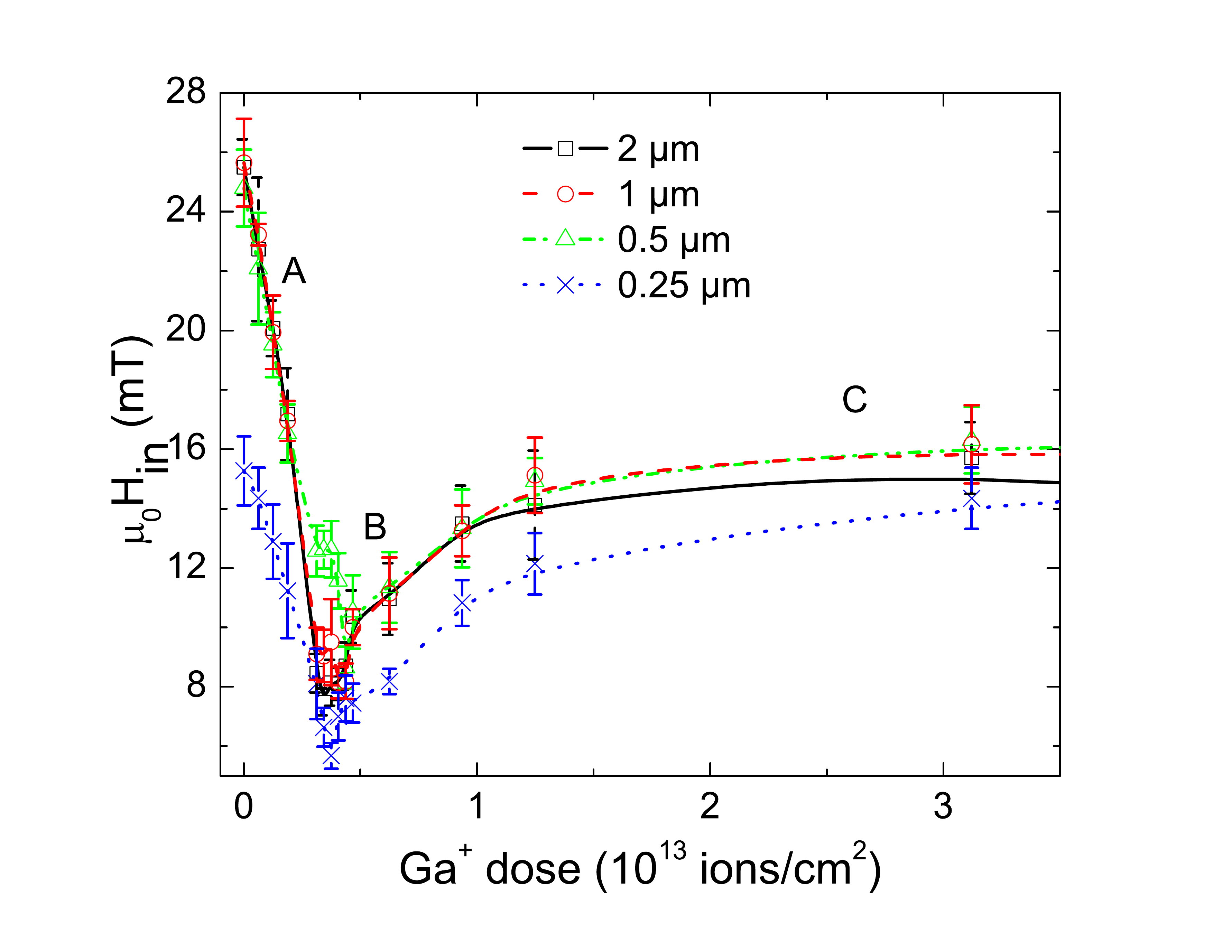}
   \caption{DW injection field as a function of Ga dose for a Pt / Co(0.6\,nm) / Pt strip of variable width. The lines are drawn as a guide to the eye.}
    \label{Figure6}
    \end{centering}
\end{figure}

First the effect of Ga irradiation is studied on strips with a fixed composition Pt(4\,nm) / Co(0.6\,nm) / Pt(2\,nm). Figure \ref{Figure5} shows exemplary Kerr images of the switching process in several 10\,$\times$1\,$\hbox{\textmu}$m$^{2}$ strips. The Kerr images of three different Ga doses are shown. In figure \ref{Figure6} the measured injection field is plotted as a function of Ga dose for structures of various sizes.

Here we discuss the features observed in the Kerr images of figure \ref{Figure5}. The samples were saturated at negative field and the field was swept to positive saturation. Snapshots at different positive fields during the sweep are shown. In figure \ref{Figure5}(a) (dose 0.34\,$\times$10$^{13}$\,ions/cm$^{2}$), it is seen that at a certain field strength, the bright structures have switched completely while the dark structures have not. This is due to the statistical nature of domain nucleation in perpendicular materials, which occurs at random defects. At a slightly higher field (figure \ref{Figure5}(b)), 2 more structures have switched instantly. This means that a DW was nucleated in the irradiated area, which instantly moves into the remainder of the strip. In other words, the nucleation field is much higher than the pinning field, $H_{\rm{n}} > H_{\rm{pin}}$. The range of doses where this is the case is denoted by A in figure \ref{Figure6}. Clearly, $H_{\rm{n}}$ decreases with Ga dose due to the PMA reduction.

In the snapshots taken at higher dose (0.41\,$\times$10$^{13}$\,ions/cm$^{2}$) in figure \ref{Figure5}(c), it is seen that a DW nucleated in the irradiated area pins at the boundary between the two regions in some strips. However, in other structures the DW moved instantly without pinning. This indicates that the field strengths associated with nucleation and pinning are approximately the same, $H_{\rm{n}} \approx H_{\rm{pin}}$. A significantly higher field is needed (figure \ref{Figure5}(d)) to depin all the trapped DWs.

Looking at a slightly higher dose of 0.44\,$\times$10$^{13}$\,ions/cm$^{2}$ in figure \ref{Figure5}(e), a strong change in the nucleation of the DW is observed. Instead of the instantaneous switching that was observed before, the irradiated area now switches in many small domains, because we are getting close to the in-plane transition. By increasing the field as seen in figure \ref{Figure5}(f), a single domain will again appear and the corresponding DW is pinned for all structures at the shown field. Hence, $H_{\rm{n}} < H_{\rm{pin}}$. This regime is denoted B in figure \ref{Figure6}.

In figure \ref{Figure6}, $H_{\rm{in}}$ as a function of Ga dose is plotted for structures of different sizes. Next to the discussed regimes A ($H_{\rm{n}}>H_{\rm{pin}}$), B ($H_{\rm{n}}<H_{\rm{pin}}$), we identify a third regime C where the pinning field converges to an asymptote, because the magnetization of the irradiated region becomes in-plane. The same 3 regimes were found in the micromagnetic model depicted in figure \ref{Figure2}. For the strips of 15$\times$2\,$\hbox{\textmu}$m$^{2}$, 10$\times$1\,$\hbox{\textmu}$m$^{2}$ and 5$\times$0.5\,$\hbox{\textmu}$m$^{2}$ the behavior is very similar. The 2.5\,$\times$0.25\,$\hbox{\textmu}$m$^{2}$ structures however behave somewhat differently. Although all the observed features are still present, it can be seen that these structures have a significantly lower nucleation field in regime A. Since all structures are grown and measured under the same conditions on the same wafer, this effect must be related to the decrease in size. Indeed, due to the limitations of the lithography method used, the roughness of the strips is very significant compared to the strip width, resulting in a rather poorly defined strip. The nucleation field is very sensitive to structural defects and is therefore reduced, and also the anisotropy itself might be affected, leading to a change of the observed effects.

The magnitude of the injection fields is roughly a factor 20 higher in the simulations/1D model compared to the experiments. This is not unusual, since the simulations do not include any thermal fluctuations. In room temperature experiments, thermal fluctuations play a crucial role in all magnetization reversal phenomena. For examle, the coercive field (responsible for the injection field in the high-$K$ range) is greatly reduced at finite temperatures, and originates from the nucleation of a small area followed by DW motion, instead of the Stoner-Wohlfart type of switching in our model. In SQUID measurements, it was found that for a similar film, the coercivity at 5\,K is roughly 40 times larger than at room temperature. Also, the escape of a DW over an energy barrier (responsible for the DW injection in the low-$K$ region) is much easier at elevated temperatures, so lower fields are required for depinning. Therefore, only a qualitative comparison with the micromagnetic model can be made.

\subsection{Variable Co layer thickness}

\begin{figure}[htb]
   \begin{centering}
   \includegraphics[width=0.7\linewidth]{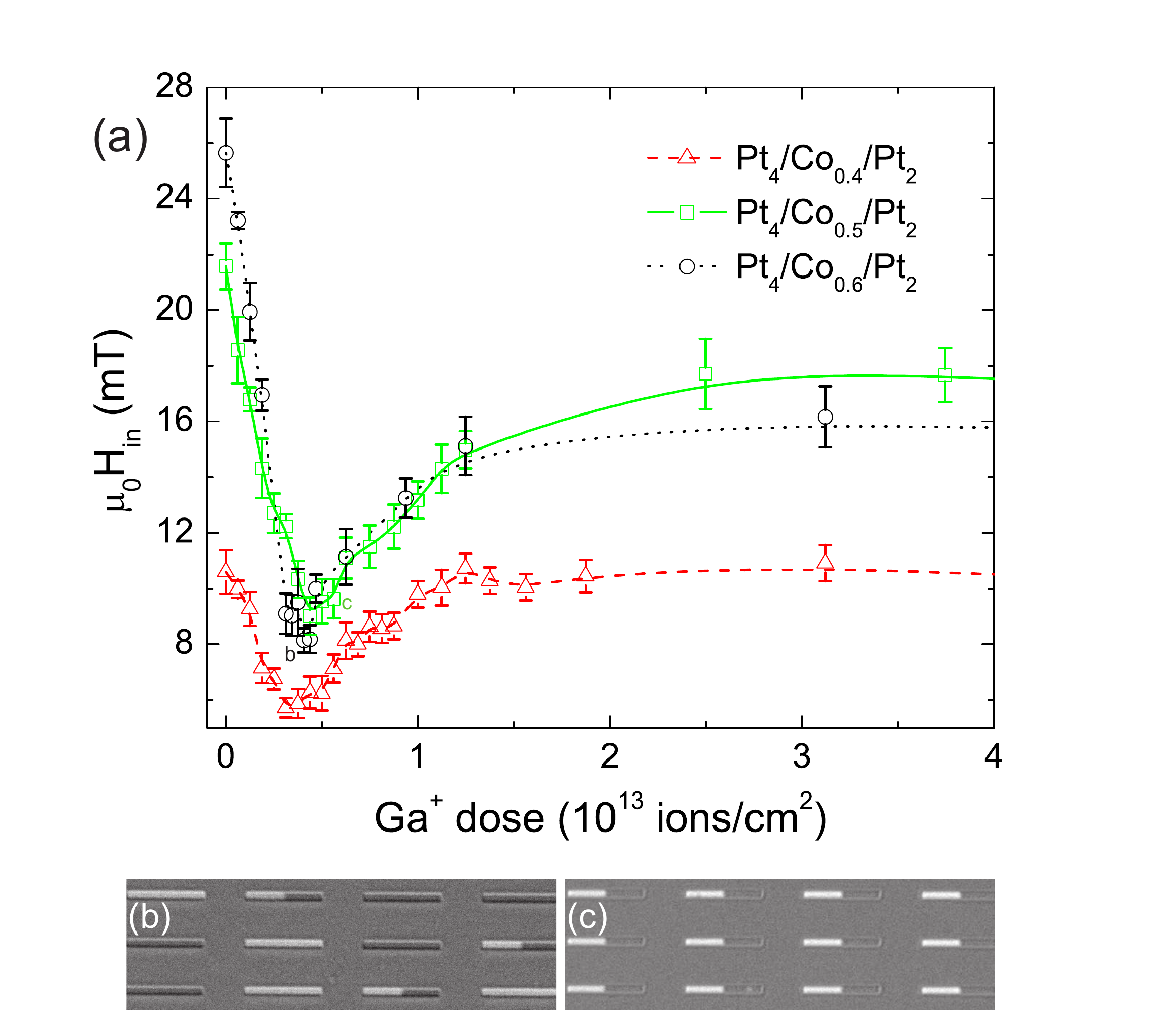}{}
   \caption{(a) DW injection field in 1\,$\hbox{\textmu}$m wide strips as a function of Ga dose for different Co thicknesses. Kerr snapshots of (b) 0.6\,nm  and (c) 0.5\,nm structures at the highest dose with full PMA, demonstrating that pinning is better tunable in a thinner Co layer.}
    \label{Figure7}
    \end{centering}
\end{figure}

Figure~\ref{Figure7} shows a comparison of $H_{\rm{in}}$ as a function of Ga dose for different Co thicknesses in Pt / Co ($x$\,nm) /Pt structures of 10\,$\times$1\,$\hbox{\textmu}$m$^{2}$. The $x=0.4\,$nm structures clearly have a lower nucleation field. This is probably related to the growth quality of such ultrathin films. Interestingly, the pinning strength is very similar for the 0.5 and 0.6\,nm Co thicknesses. This is also what would be expected from the anisotropy measurements of figure \ref{Figure4}, because $K_{\rm{eff,0}}-K_{\rm{eff}}$ appeared to be rather insensitive to the layer thickness. The minimum of the curve, where $H_{\rm{pin}}=H_{\rm{n}}$, is found at a dose of 0.44\,$\times$10$^{13}$ions/cm$^{2}$ for both the 0.5\,nm and 0.6\,nm strips. For the 0.4\,nm structures $H_{\rm{n}}$ is lower (related to the growth quality of such thin layers), which shifts the minimum slightly to the left at 0.31\,$\times$10$^{13}$ions/cm$^{2}$. Also, the DW pinning in regime B is lower for the 0.4\,nm strips, because the anisotropy is better retained at high doses compared to the 0.5\,nm sample (as seen in figure \ref{Figure4}), leading to a lower pinning barrier. In the high-dose regime (C), where the irradiated region has an in-plane magnetization, $H_{\rm{pin}}$ is theoretically given by $K_{\rm{eff,0}}/(\mu_0 M_{\rm{s}})$, so ultimately determined by the anisotropy of the untouched part $K_{\rm{eff,0}}$. Both $K_{\rm{eff,0}}$ and  $H_{\rm{pin}}$ are significantly higher for the 0.5\,nm Co film, demonstrating that the theoretical model appears to have qualitative validity also in this regime. For the 0.4\,nm Co film, the pinning field at high dose is masked by the very low $H_{\rm{n}}$.

Compatible with the anisotropy measurements in figure \ref{Figure4}, it is seen from the Kerr images that for thin Co layers, much larger anisotropy differences can be obtained before the magnetization becomes in-plane. Because theoretically $H_{\rm{pin}}=(K_{\rm{eff,0}}-K_{\rm{eff}})/(\mu_{0}M_{\rm{s}})$ this means that the pinning strength of the anisotropy barrier can also be made much stronger. Decreasing the Co thickness therefore leads to more controllable DW pinning. This is illustrated by figure \ref{Figure7}(c), which shows that DWs are consistently pinned in the 0.5\,nm Co strip for all the studied structures at the shown dose of $0.56\times10^{13}\,$ions/cm$^2$. At the same dose, the 0.6\,nm Co strip is already in-plane magnetized. The highest dose where the 0.6\,nm strips are fully perpendicular is $0.41\times10^{13}\,$ions/cm$^2$, and figure \ref{Figure7}(c) illustrates the unreliable pinning in these strips. For application as pinning sites, one typically would like to pin an existing domain wall without risking nucleation of a new domain wall. Therefore, one would require a significant gap between the highest $H_{\rm{n}}$ and the lowest $H_{\rm{pin}}$ of any of the structures. For the 0.6\,nm Co, this gap is virtually zero for any dose with full PMA. For 0.5\,nm, the gap is maximized at $0.56\times10^{13}\,$ions/cm$^2$ and 0.8 mT in size. Interestingly, for the 0.4\,nm strips, full PMA extends to very high doses and the optimal gap was 4.7\,mT at a dose of $0.81\times10^{13}\,$ions/cm$^2$.

\subsection{Tuning the width of the pinning barrier}

In the previous sections we showed that the DW pinning field at a Ga irradiation boundary scales with $K_{\rm{eff,0}}-K_{\rm{eff}}$, where $K_{\rm{eff,0}}$ can be tuned by the Co interlayer thickness and $K_{\rm{eff}}$ by the Ga dose. However, equation \ref{e:gallium:pinningfield} suggests another parameter to tune the pinning field: the length scale of the anisotropy gradient $\delta$. It is expected that the pinning strength decreases with increasing $\delta$, because the energy barrier for DW propagation becomes less steep. Experimentally, $\delta$ is controlled by placing the sample away from the focal point. The distance to the focal point determines the FWHM of the beam, which is used as an estimate of $\delta$.

\begin{figure}[htb]
   \begin{centering}
   \includegraphics[width=0.7\linewidth]{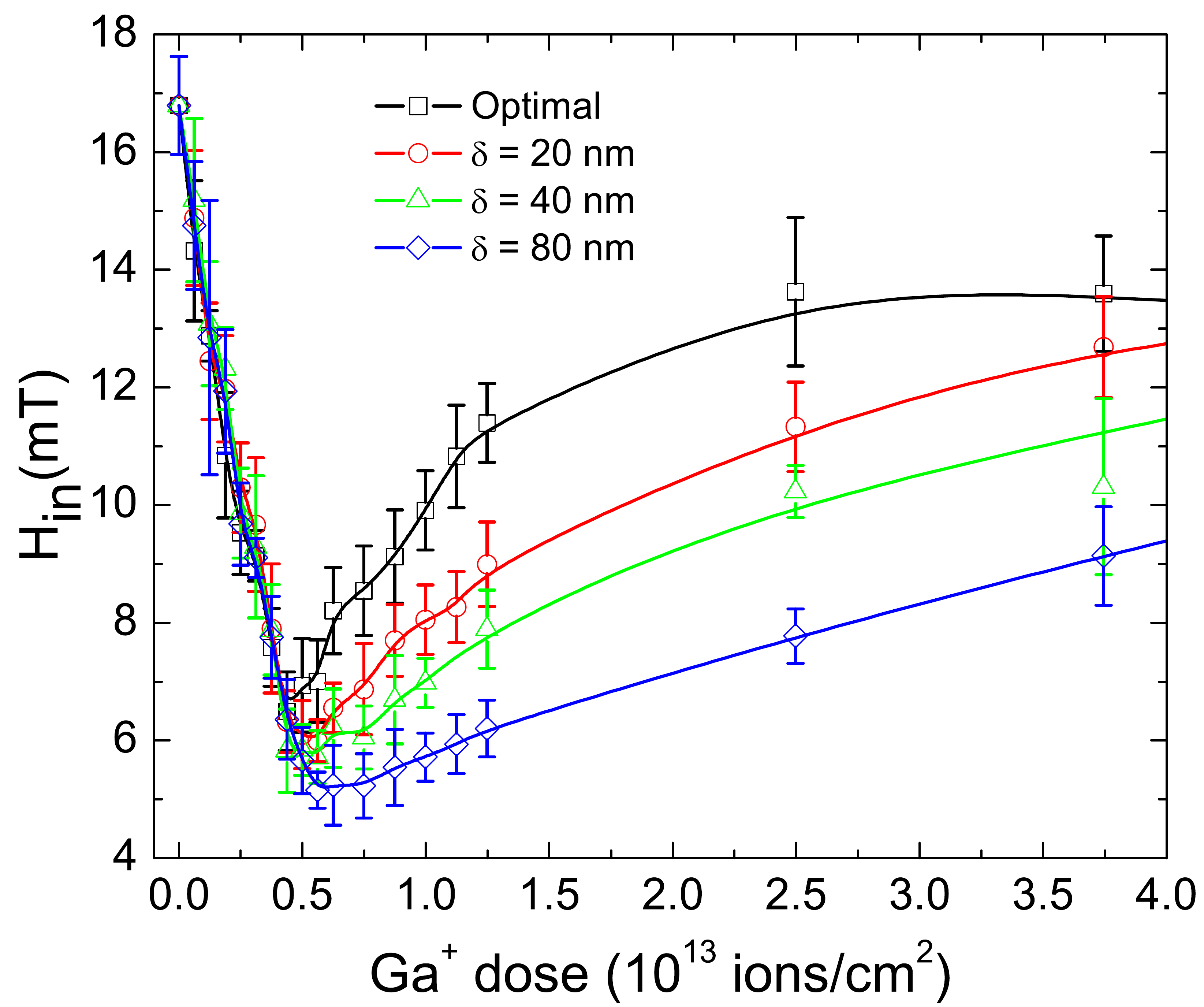}{}
   \caption{DW injection field of 1\,$\hbox{\textmu}$m wide Pt / Co(0.5\,nm) / Pt structures. The width of the anisotropy barrier $\delta$ is controlled by changing the focus of the ion beam. As expected, the pinning strength is reduced for increasing $\delta$.}
    \label{Figure8}
    \end{centering}
\end{figure}

Figure~\ref{Figure8} illustrates the behavior of the injection field in Pt / Co (0.5\,nm) / Pt as $\delta$ is varied from 0 (optimal beam focus) to  $\approx 80$\,nm. Increasing $\delta$ clearly leads to a systematic decrease of $H_{\rm{pin}}$. The qualitative agreement with the theoretical result of figure \ref{Figure2} is striking. The fact that a slight change of $\delta$ leads to such clear effects is strong evidence that Ga irradiation creates pinning sites at a length scale comparable to the DW width. Using focused helium beams, an even smaller $\delta$ can be realized due to a better optimal focus, leading to stronger DW pinning \cite{Markie}. It is interesting to note that the minimum in $H_{\rm{in}}$ is also reduced when increasing $\delta$. A lesson to learn from this, is that in order to achieve DW injection at the lowest possible field, one should simply make $\delta$ as big as possible.

\section{Conclusion}

In this paper, we analyzed in detail the pinning of a domain wall at engineered anisotropy variations. First, we analytically derived that a step in the magnetic anisotropy acts as an energy barrier for the DW. It was shown that the pinning field of a DW at such an anisotropy boundary increases with the anisotropy difference and decreases with the width of the boundary. The analytical model matches well with micromagnetic simulations. Then, it was shown that FIB irradiation with Ga ions can be used to control the magnetic anisotropy of a Pt/Co/Pt strip, and quantitative measurements were performed using the EHE effect. Thereafter, field-induced domain wall pinning and nucleation in irradiated Pt/Co/Pt nanostrips was studied using wide-field Kerr microscopy. The pinning behavior qualitatively reproduced all the features of the analytical model. The pinning of DWs was shown to be insensitive to the width of the strip in the range 0.5-2\,$\hbox{\textmu}$m. However, the thickness of the Co layer does provide another handle to tune DW pinning, since a thinner Co layer has higher intrinsic anisotropy, thereby increasing the range of anisotropy values that can be realized without destroying the PMA. Finally, it was shown that even the width of the anisotropy barrier, which according to our model has to be of the order of the DW width ($\sim10\,$nm), can be precisely tuned by reducing the focus of the ion beam. This leads to a lower injection field because the energy barrier for the DW becomes less steep.

Engineered anisotropy defects can not only be used to controllably inject a DW at arbitrarily low fields, but also to provide tunable pinning sites for field- and current-induced domain wall motion in PMA strips. In the experiments reported in this paper, relatively large areas were irradiated with Ga, but also small defects could be made that act as pinning sites. These can be useful in DW-based memory or logic devices as an alternative to geometrically induced pinning sites \cite{Parkin2008,Zutic2004,Xu2008}, or for controlled experiments on current-induced DW depinning. Furthermore, we have recently shown by micromagnetic simulations that a DW pinned at an anisotropy boundary can be brought into steady oscillatory motion by a DC current \cite{oscillator}, which could be used as a microwave current source similar to spin torque oscillators. To conclude, control of the magnetic anisotropy at the nanoscale in general is a powerful tool in many magnetic nanodevices.

\section{Acknowledgement}
This work is part of the research programme of the Foundation for Fundamental Research on Matter (FOM), which is part of the Netherlands Organisation for Scientific Research (NWO).

This is an author-created, un-copyedited version of an article accepted for publication in Journal of Physics: Condensed Matter. IOP Publishing Ltd is not responsible for any errors or omissions in this version of the manuscript or any version derived from it. The definitive publisher-authenticated version is available online at http://dx.doi.org/10.1088/0953-8984/24/2/024216

\section*{References}
\bibliographystyle{jphys}
\bibliography{References}

\end{document}